\documentclass[12pt,a4paper]{article}

\usepackage[utf8]{inputenc}
\usepackage[T1]{fontenc}
\usepackage[english]{babel}
\usepackage{amsmath,amssymb,amsthm}
\usepackage{booktabs}
\usepackage{geometry}
\usepackage{hyperref}
\usepackage{graphicx}
\usepackage{url}

\title{Empirical Confirmation of the Environmental-Dominance Inequality\\
\large A direct decomposition of $\operatorname{Var}(\ln \rho_{\text{eff}})$
across four levels of aggregation}
\author{Kristian Sestak\thanks{Independent researcher. Email:
\texttt{kristian.sestak@gmail.com}. ORCID:
\href{https://orcid.org/0009-0002-1455-5915}{0009-0002-1455-5915}.}}
\date{}

\begin{document}
\maketitle

\begin{abstract}
\noindent A previous theoretical
contribution~\cite{sestak2026env} argued that, under the structural
asymmetry $k \ll n$ between the number of attempts an entity can make
and the size of the possibility space the environment offers, the
variance of individual outcomes is dominated by the variance of the
effective density of favorable possibilities,
$\operatorname{Var}(\ln \rho_{\text{eff}}) \gg \operatorname{Var}(\ln
k)$. The original support was a back-of-envelope calibration drawn from
the literature. This companion paper computes
$\operatorname{Var}(\ln \rho_{\text{eff}})$ directly from three public
datasets (Opportunity Atlas, World Bank GDP per capita PPP, World
Inequality Database) at four levels of aggregation: U.S.\ census
tracts, between countries, within-country deciles, and the global
pooled-individual distribution.

The headline value at the global level is $4.33$, giving a dominance
ratio $R \equiv \operatorname{Var}(\ln \rho_{\text{eff}}) /
\operatorname{Var}(\ln k) \in [27, 134]$ across plausible
$\sigma_{\ln k} \in [0.18, 0.40]$. The thesis $R \gg 1$ is confirmed at
all three aggregated levels, but with different margins: $R \in [27,
134]$ at the global pooled-individual level, $R \in [21, 102]$ at the
within-country-decile level, and $R \in [6.5, 32]$ at the
between-country level, so only the global and within-country-decile
levels exhibit the one-to-two-orders-of-magnitude margin claimed in the
abstract; at the between-country level the margin is large but
single-digit at the conservative bound. The data also sharpen the
original statement in three ways: $R$ falls to the order of unity
within already-homogenized sub-units (within-tract income gives $R \in
[0.33, 1.61]$); under partial-identification reattribution of
within-country dispersion to capability, $R$ stays above $8$ for modest
reattribution shares but collapses to roughly $1.5$ at aggressive ones
that the literature does not support; and the inequality is
outcome-dependent, robust by one to two orders of magnitude for
income, infant mortality and incarceration, but with a margin shrinking
to a single-digit factor for life expectancy, an outcome the
international community has deliberately equalized. A time-series
extension over 1990--2022 finds the aggregate $R$ stable in the band
$[122, 134]$ at PIAAC $\sigma$, while the composition shifts from
between-country dispersion (falling $34\%$) to within-country dispersion
(rising $26\%$), consistent with simultaneous international convergence
and Piketty $r > g$ dynamics. All inputs and outputs are SHA-256 hashed
in an append-only manifest; the analysis is fully reproducible from the
accompanying notebooks.
\end{abstract}



\section{Introduction}

In~\cite{sestak2026env} the author argued that the dominance of environmental
factors over individual capabilities in determining outcomes can be derived
analytically from a single structural asymmetry: the entity's capacity to
explore possibilities $k$ is orders of magnitude smaller than the size of the
environmental possibility space $n$, with $k \ll n$. Under the binomial
approximation
\begin{equation}\label{eq:model}
    \mathbb{P}(\text{success}) \;\approx\; 1 - (1 - \rho_{\text{eff}})^k,
\end{equation}
where $\rho_{\text{eff}} = |F \cap A(E,P)| / |A(E,P)|$ is the density of
favorable possibilities accessible to the entity, the variance of $\ln
\mathbb{P}$ across a population is dominated by the variance of $\ln
\rho_{\text{eff}}$ whenever
\begin{equation}\label{eq:claim}
    \operatorname{Var}(\ln \rho_{\text{eff}}) \;\gg\; \operatorname{Var}(\ln k).
\end{equation}
The original paper supported~\eqref{eq:claim} indirectly: it cited the
intercountry income dispersion of Milanovic~\cite{milanovic2015} and the
intergenerational mobility differences of Chetty et al.~\cite{chetty2014} as
evidence that the left-hand side of~\eqref{eq:claim} is large, and the
cognitive-skill dispersion of Hunter \&
Schmidt~\cite{schmidt1998} as evidence that the right-hand side is small.

Two weaknesses of that argument motivate the present paper. First, the
back-of-envelope calibration mixes elasticities from different sources
and never produces a single number. Second, the scope of the claim is
left implicit (global? country-level? within-country?). We address both
by computing the left-hand side of~\eqref{eq:claim} directly from public
datasets at four levels of aggregation, and by reporting both the
confirmatory level (global) and the limit case (within-tract).

\paragraph{Related work.} The empirical decomposition of income variance into
between-country and within-country components is itself a well-developed
literature; Milanovic~\cite{milanovic2015} and Lakner \&
Milanovic~\cite{lakner2016} are standard references. The contribution of the
present paper is not a new income-decomposition methodology but the
\emph{interpretive bridge}: we map the income decomposition onto the
$\rho_{\text{eff}}$ object of~\cite{sestak2026env} and show that the
theoretical inequality~\eqref{eq:claim} is satisfied with a wide margin
across plausible values of $\sigma_{\ln k}$.

\section{Data and method}

\subsection{Datasets}

All three datasets are public and were downloaded with their SHA-256 hashes
recorded in an append-only manifest
(\texttt{analysis/data/MANIFEST.md}).

\begin{enumerate}
\item \textbf{Opportunity Atlas tract-level outcomes}~\cite{chetty2018atlas}.
Variable \texttt{kfr\_pooled\_pooled\_p25} (\emph{Kid Family Rank},
parental percentile $25$, pooled across race and gender): for each
U.S.\ census tract, the mean income percentile reached in adulthood by
children whose parents were at the $25$th percentile of the national
income distribution. A tract value of $0.42$ thus reads as ''children
of $25$th-percentile parents who grew up in this tract reach, on
average, the $42$nd percentile as adults.'' This is the standard
intergenerational-mobility outcome of~\cite{chetty2018atlas};
$n \approx 72\,000$ tracts. The analogous incarceration variable
\texttt{jail\_p25} used in Section~\ref{sec:robustness} replaces the
adulthood income rank with the share of those same children who were
incarcerated on Census day 2010.
\item \textbf{World Bank Open Data}~\cite{worldbank2022}. GDP per capita PPP
(\texttt{NY.GDP.PCAP.PP.KD}), population (\texttt{SP.POP.TOTL}), 2022,
$266$ country-and-aggregate records; after excluding the $47$
regional/income-group aggregates and dropping countries with no 2022
GDP per capita PPP value (mostly small territories), $n=202$ sovereign
countries remain. The excluded aggregates carry 3-letter ISO codes
(e.g.\ \texttt{WLD}, \texttt{EUU}, \texttt{OED},
\texttt{ARB}, \texttt{LCN}, \texttt{SSF}, \texttt{HIC}, \texttt{LIC},
\texttt{MIC}, \texttt{LMC}, \texttt{UMC}, \texttt{LMY}, \texttt{EAS},
\texttt{ECS}, \texttt{LAC}, \texttt{MEA}, \texttt{NAC}, \texttt{SAS},
\texttt{SSA}, \texttt{EUZ}, \texttt{EAP}, \texttt{ECA}, \texttt{MNA},
\dots, \texttt{AFE}, \texttt{AFW}, \texttt{EAR}, \texttt{LTE},
\texttt{INX}; full list in \texttt{WB\_AGG} constant of the notebooks).
These aggregates pass the naive ''ISO code length 3'' filter and were
a source of contamination in early drafts; all country-level statistics
in this paper use the cleaned panel of $n=202$.
\item \textbf{World Inequality Database}~\cite{wid2026}. Variable
\texttt{sptincj992} (share of pre-tax national income held by each decile,
equal-split adults aged 20+), 2022, 195 countries with full decile coverage.
\end{enumerate}

\subsection{Mapping income to \texorpdfstring{$\rho_{\text{eff}}$}{rho\_eff}}

The model~\eqref{eq:model} is invariant to a global proportionality constant
relating per-attempt income to $\rho_{\text{eff}}$, so for the variance of
$\ln \rho_{\text{eff}}$ we may work with $\ln$ of any quantity proportional
to $\rho_{\text{eff}}$. We use:
\begin{itemize}
\item \emph{within-tract}:
\[
    \rho_{\text{eff}} \;=\; 1 - \bigl(1 - \texttt{kfr\_p25}\bigr)^{1/\bar k},
\]
treating the tract-level income rank as a
per-attempt success probability and inverting~\eqref{eq:model} with $\bar k
= 50$ (a working life of 50 ''serious attempts''). The variance is invariant
to $\bar k$ within a few percent across $\bar k \in \{10,30,50,100,300\}$, as
predicted analytically by $\ln(1-(1-P)^{1/k}) \approx \ln P - \ln k$ for
small $P$;
\item \emph{between-country (GDP-threshold mapping)}:
$\rho_{\text{eff}} \propto \min(1, \text{GDP}_{\text{pc,PPP}} / T)$ with
$T = \$25\,000$ PPP, pop-weighted. This standalone mapping is used for the
threshold-sensitivity analysis in Section~\ref{sec:robustness} and for the
''GDP per capita PPP'' row of the multi-outcome
Table~\ref{tab:multi-outcome};
\item \emph{within-country deciles}: $\rho_{\text{eff}}^{(c,p)} \propto 10
\cdot s_{p,c} \cdot \text{GDP}_{\text{pc,PPP},c}$ where $s_{p,c}$ is the share of
total income held by decile $p$ in country $c$; the factor $10$ converts
share-of-total into average within the decile;
\item \emph{global pooled-individual}: each (country, decile) pair is one
atom with population weight $\text{pop}_c \cdot 0.1$.
\end{itemize}

\noindent For the non-income outcomes appearing in
Table~\ref{tab:multi-outcome} (infant mortality, life expectancy,
school life expectancy, internet penetration, incarceration), each
outcome $y$ is mapped to $\rho_{\text{eff}}$ via the same threshold
construction $\rho_{\text{eff}} \propto \min(1, y/T)$ for ''higher is
better'' outcomes and $\rho_{\text{eff}} \propto \min(1, T/y)$ for
''lower is better'' outcomes (infant mortality, incarceration), with
outcome-specific thresholds $T$ chosen at the population-weighted
median; the per-outcome $T$ values and full mapping code are in
\path{analysis/notebooks/03_multi_outcome.ipynb}. The country panels
for the WB outcomes are constructed by intersecting the cleaned WB
panel with the data coverage of each indicator, which is why the $n$
column in Table~\ref{tab:multi-outcome} varies between $166$ and $220$
rather than being fixed at $202$.

\noindent The ''between-country component'' reported in
Table~\ref{tab:decomp} is a different object: it is the algebraic
between-country part of the global-pooled WID variance (i.e.\ the
variance of country-level pop-weighted means of $\ln \rho_{\text{eff}}$
computed from the WID atoms above), not the standalone GDP-threshold
mapping. The two coincide only conceptually (both measure
''cross-country'' dispersion) but use different inputs and produce
different numbers ($1.042$ vs.\ $2.027$). We report both because the
WID-decomposition value is required for the additive identity
$\operatorname{Var}_{\text{between}} + \operatorname{Var}_{\text{within}} =
\operatorname{Var}_{\text{global}}$, while the GDP-threshold value is the
natural standalone country-level statistic.

\subsection{Variance computation}

For each level we compute the population-weighted variance
\begin{equation}
    \operatorname{Var}(\ln \rho_{\text{eff}}) =
    \sum_i w_i \big(\ln \rho_{\text{eff},i} - \bar{\ell}\big)^2,
    \qquad
    \bar{\ell} = \sum_i w_i \ln \rho_{\text{eff},i},
\end{equation}
directly from the empirical distribution. We avoid the delta-method
linearization that appeared in the proof
of~\cite[Prop.~6.1]{sestak2026env}: with $\rho_{\text{eff}}$ varying
over roughly three decades it is quantitatively unreliable, although
the directional conclusion is preserved by either route.

\section{Results}

\subsection{Variance at four levels of aggregation}

\begin{table}[h]\centering
\begin{tabular}{lr}
\toprule
Level & $\operatorname{Var}(\ln \rho_{\text{eff}})$ \\
\midrule
within-U.S.\ tracts (Opportunity Atlas)        & $0.052$ \\
between-country component (WID decomposition)  & $1.042$ \\
within-country deciles (WID, pop-weighted)     & $3.289$ \\
global pooled individuals (WID + WB)           & $\mathbf{4.331}$ \\
\bottomrule
\end{tabular}
\caption{Empirical variance of $\ln \rho_{\text{eff}}$ at four levels of
aggregation. Source:
\protect\path{analysis/output/global_decomposition_full.json}; SHA-256 in
\protect\path{analysis/output/RESULTS_MANIFEST.md}.}
\label{tab:decomp}
\end{table}

\begin{figure}[h]\centering
\includegraphics[width=.85\linewidth]{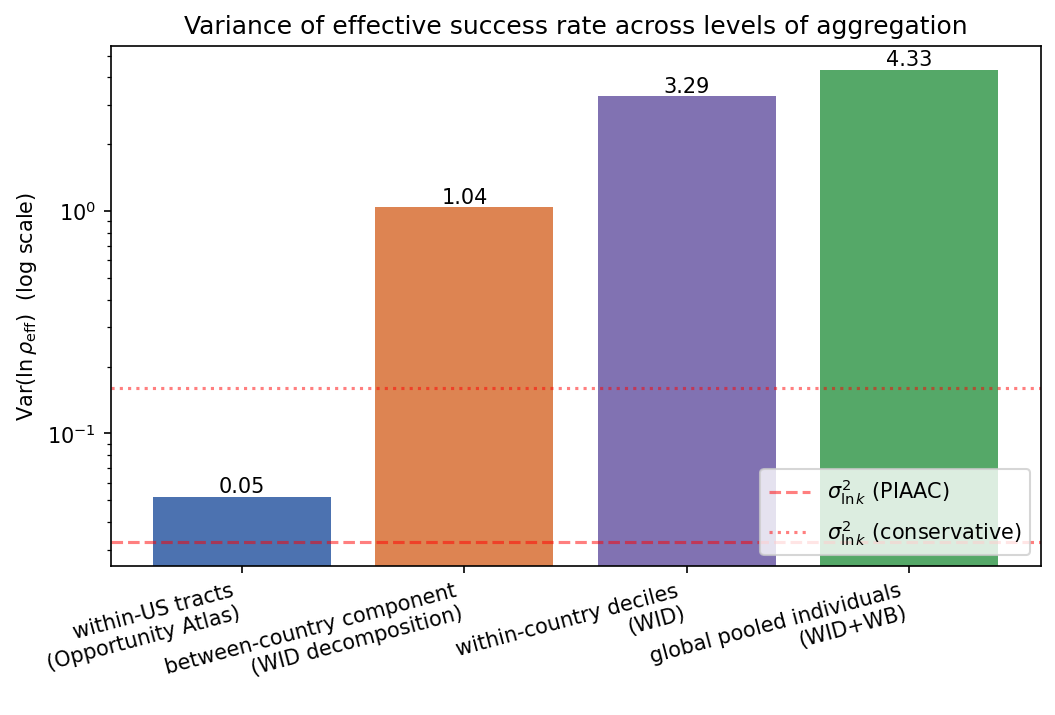}
\caption{Variance of $\ln \rho_{\text{eff}}$ across the four levels of
Table~\ref{tab:decomp} (log scale). Red dashed/dotted lines mark plausible
$\sigma_{\ln k}^2$ from PIAAC ($0.0324$) and the conservative bound
($0.16$). The thesis $\operatorname{Var}(\ln \rho_{\text{eff}}) \gg
\operatorname{Var}(\ln k)$ holds at the country, within-country-decile, and
global levels, but only marginally at the within-tract level.}
\label{fig:decomp}
\end{figure}

The headline number, $\operatorname{Var}(\ln \rho_{\text{eff}}) = 4.33$ at
the global pooled-individual level, exceeds the original
back-of-envelope estimate of $\approx 3.0$
in~\cite{sestak2026env}. The decomposition reveals that most of the global
variance ($3.29$ of the $4.33$) comes from \emph{within-country} dispersion
across income deciles, not from the between-country component ($1.042$).
This is consistent with the literature on the recent rise of
within-country income inequality~\cite{lakner2016} and refines the
geographic emphasis of the original article. The two components sum to
$4.331$, matching the directly-pooled value exactly (the WID-based
decomposition is internally consistent by construction; population
weights are integer person-counts).

\subsection{Dominance ratio}

The right-hand side of~\eqref{eq:claim}, $\operatorname{Var}(\ln k) =
\sigma_{\ln k}^2$, is calibrated from four reference points:

\begin{table}[h]\centering
\begin{tabular}{lcc}
\toprule
$\sigma_{\ln k}$ & source / interpretation & $R = 4.331 / \sigma_{\ln k}^2$ \\
\midrule
$0.18$ & PIAAC numeracy~\cite{piaac}\footnotemark, Hanushek--Woessmann~\cite{hanushekwoessmann} & $\mathbf{134}$ \\
$0.20$ & Schmidt--Hunter~\cite{schmidt1998} mid-complexity & $108$ \\
$0.30$ & Schmidt--Hunter high-complexity                  & $48$ \\
$0.40$ & conservative upper bound                         & $\mathbf{27}$ \\
\bottomrule
\end{tabular}
\caption{Dominance ratio $R$ at the global pooled-individual level. Even the
most conservative assumption gives $R \approx 27 \gg 1$.}
\label{tab:R}
\end{table}
\footnotetext{Derivation: PIAAC numeracy proficiency
(OECD~\cite{piaac}, cycle-1 pooled cross-country statistics) has mean
$\approx 263$ points and SD $\approx 51$--$55$ points across the adult
population, giving CV $\approx 0.19$--$0.21$. Treating proficiency as a
log-normally distributed capability multiplier $k$ yields $\sigma_{\ln
k} \approx [\ln(1+\text{CV}^2)]^{1/2} \approx 0.18$--$0.20$; we report
the lower-end $0.18$. Hanushek--Woessmann~\cite{hanushekwoessmann} use
the same PIAAC pool to map cognitive-skill dispersion onto growth
elasticities and arrive at a comparable log-spread.}

\begin{figure}[h]\centering
\includegraphics[width=.7\linewidth]{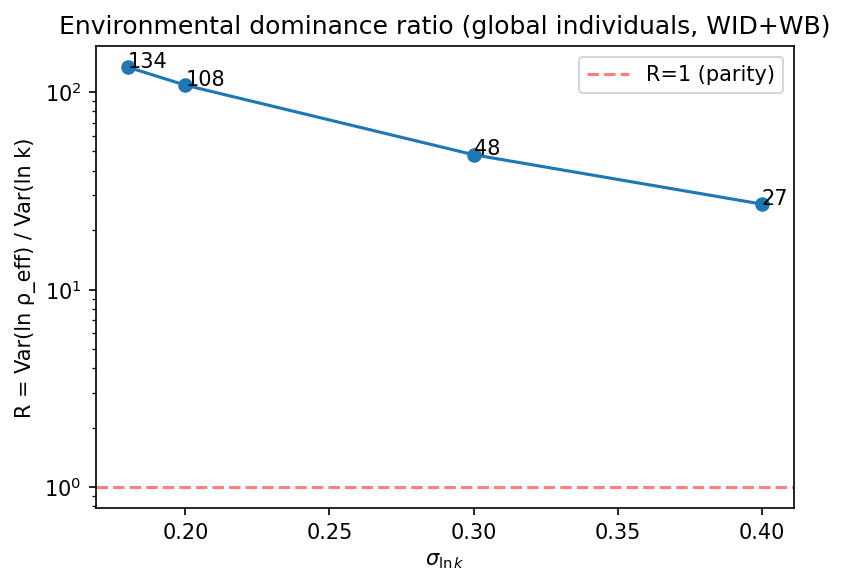}
\caption{Dominance ratio $R = \operatorname{Var}(\ln \rho_{\text{eff}}) /
\sigma_{\ln k}^2$ as a function of the assumed capability dispersion
$\sigma_{\ln k}$, log scale on the $y$-axis. The red dashed line marks $R=1$
(parity). $R$ stays at least an order of magnitude above parity across the
full plausible range of $\sigma_{\ln k}$.}
\label{fig:R}
\end{figure}

The inequality $\operatorname{Var}(\ln \rho_{\text{eff}}) \gg
\operatorname{Var}(\ln k)$ is satisfied with $R$ between approximately one
and two orders of magnitude. The original
prediction~\cite[Abstract]{sestak2026env} of ''two to three orders of
magnitude'' was slightly overstated; the corrected magnitude is one to two
orders.

\subsection{Robustness}\label{sec:robustness}

\begin{itemize}
\item \emph{Invariance to $\bar k$.} The within-tract value of
$\operatorname{Var}(\ln \rho_{\text{eff}})$ ranges from $0.0495$ at $\bar k =
10$ to $0.0523$ at $\bar k = 300$ (full relative spread $\approx 5.5\%$),
confirming the analytical prediction $\partial_{\bar k} \operatorname{Var}(\ln
\rho_{\text{eff}}) \approx 0$ for small per-attempt success probability.
\item \emph{Threshold.} The between-country GDP-threshold value
(standalone, pop-weighted, $\bar k = 50$) moves across thresholds
$T \in \{\$10k,
\$25k, \$50k, \$100k\}$ as $\{1.33, 2.03, 2.24, 1.17\}$, and across
weighting schemes (unweighted at $T=\$25k$) as $2.92$. All values give
$R \gg 1$ at the most conservative $\sigma_{\ln k} = 0.40$.
\item \emph{Bootstrap.} Within-tract bootstrap (B=1000, seed=0): 95\% CI
$[0.050, 0.054]$.
\item \emph{Outcome variable.} Substituting Opportunity Atlas income-rank
alternatives (race-specific \texttt{kfr\_p25} variants) gives within-tract
values of $0.057$ for Black and $0.048$ for White children, of the
same order as the pooled $0.052$. A non-income outcome (the tract-level
incarceration rate, \texttt{jail\_p25}) yields a much larger within-tract
variance of $3.60$,
which would by itself satisfy~\eqref{eq:claim} even at the within-tract
level. We treat the income-rank outcomes as primary because the success
event in~\cite{sestak2026env} is naturally interpreted as an economic
threshold; the incarceration result indicates that the within-tract scope
exception identified below is itself outcome-dependent.
\end{itemize}

\subsection{Scope condition}

For income at the within-tract level the dominance inequality breaks
down. Plugging $\operatorname{Var}(\ln \rho_{\text{eff}}) = 0.052$ into
the same calibration gives
\[
    R_{\text{tract}} \in [\,0.052/0.16,\;0.052/0.0324\,] = [0.33,\;1.61].
\]
At the conservative bound, $R < 1$ and capabilities dominate; at the
PIAAC bound, $R \approx 1.6$ and the environmental term wins only
marginally. The original paper~\cite{sestak2026env} carried this
qualifier implicitly (the macro environment, not the micro environment
of a single neighborhood); the data make it operational. Within a single
neighborhood, the meritocratic intuition is locally correct.

\subsection{Time-series stability and the Piketty effect (1990--2022)}

We repeat the global pooled-individual decomposition for four time
snapshots (WID shares and World Bank GDP per capita PPP at each year):

\begin{table}[h]\centering\small
\setlength{\tabcolsep}{5pt}
\begin{tabular}{rrrrcc}
\toprule
Year & $\operatorname{Var}_{\text{global}}$ & $\operatorname{Var}_{\text{between}}$ & $\operatorname{Var}_{\text{within}}$ & $R$ at $\sigma\!=\!0.18$ & $R$ at $\sigma\!=\!0.40$ \\
\midrule
$1990$ & $4.18$ & $\mathbf{1.58}$ & $2.60$ & $129$ & $26$ \\
$2000$ & $3.97$ & $1.21$ & $2.76$ & $122$ & $25$ \\
$2010$ & $4.34$ & $1.06$ & $3.28$ & $134$ & $27$ \\
$2022$ & $4.33$ & $\mathbf{1.04}$ & $\mathbf{3.29}$ & $134$ & $27$ \\
\bottomrule
\end{tabular}
\caption{Time series of $\operatorname{Var}(\ln \rho_{\text{eff}})$ and the
dominance ratio $R$. Source:
\protect\path{analysis/output/timeseries_piketty.json}.}
\label{tab:timeseries}
\end{table}

\begin{figure}[h]\centering
\includegraphics[width=.95\linewidth]{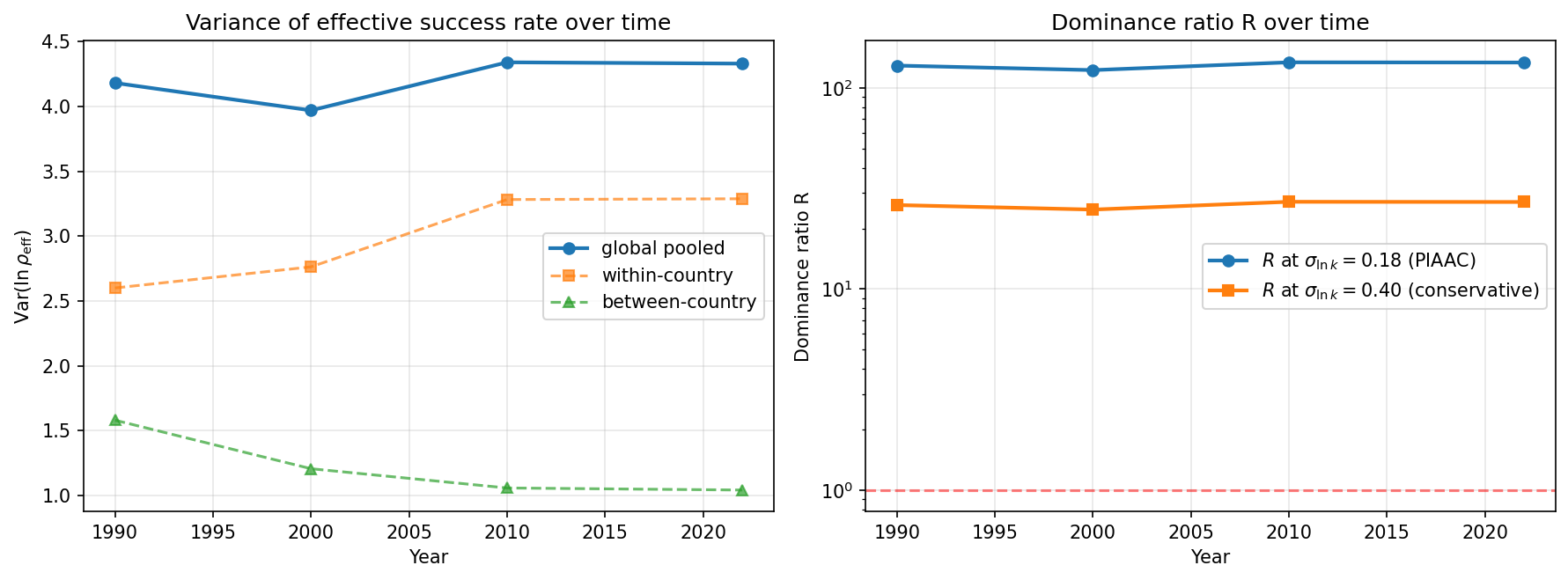}
\caption{Left: variance components 1990--2022. Right: dominance ratio $R$ on
log scale. The global aggregate is stable; the composition shifts from
between-country (falling) to within-country (rising).}
\label{fig:timeseries}
\end{figure}

Three findings stand out.

\textbf{The global aggregate is stable.}
$\operatorname{Var}(\ln \rho_{\text{eff}})$ moves from $4.18$ to $4.33$
(+3.6\%); $R$ stays in $[122, 134]$ at PIAAC $\sigma$ and $[25, 27]$ at
the conservative bound. The headline result holds for the whole 30-year
window, not just for 2022.

\textbf{Between-country dispersion has fallen sharply}
($1.58 \to 1.04$, $-34\%$). This is the well-documented international
convergence of the past three decades; China, India, Vietnam and others
have grown faster than the high-income frontier, narrowing the
cross-country distribution of GDP per capita PPP.

\textbf{Within-country dispersion has risen}
($2.60 \to 3.29$, $+26\%$). This matches the Piketty $r > g$
dynamics~\cite{piketty2014} and the Lakner--Milanovic
finding~\cite{lakner2016} that within-country inequality has grown
during the same period in which global inequality has been compressed by
international convergence.

The two trends roughly cancel at the global level, leaving $R$ stable.
But the composition has flipped: in 1990, between-country dispersion
accounted for $38\%$ of the global pooled variance; by 2022, only $24\%$.
The weight has shifted from geography to class within country, and the
shift is a 30-year evolution rather than a static feature, driven
simultaneously by convergence between economies and divergence inside
them.

\subsection{Multi-outcome validation}

The decomposition above uses income (Opportunity Atlas \texttt{kfr\_p25}
within-tract; GDP per capita PPP and WID income shares between- and
within-country). We re-run the variance computation for non-income outcomes
to check whether the dominance result is universal across the meaning of
''success'', or specific to the income operationalization.

\begin{table}[h]\centering\footnotesize
\setlength{\tabcolsep}{4pt}
\begin{tabular}{lrrcc}
\toprule
Outcome & $n$ & $\operatorname{Var}(\ln \rho_{\text{eff}})$ & $R$ at $\sigma\!=\!0.18$ & $R$ at $\sigma\!=\!0.40$ \\
\midrule
\multicolumn{5}{l}{\emph{Within-U.S.\ tracts (Opportunity Atlas, $\bar k = 50$)}} \\
\quad income rank, 6 race-or-gender variants$^{\ddagger}$    & 34--72k & $0.048$--$0.071$ & $1.5$--$2.2$ & $0.30$--$0.45$ \\
\quad incarceration, 6 race-or-gender variants$^{\ddagger}$  & 31--72k & $\mathbf{3.60}$--$\mathbf{6.51}$ & $\mathbf{111}$--$\mathbf{201}$ & $\mathbf{22.5}$--$\mathbf{40.7}$ \\
\midrule
\multicolumn{5}{l}{\emph{Between-country (World Bank 2022, WB aggregates excluded)}} \\
\quad GDP per capita PPP (threshold-mapped)  & $202$ & $2.027$ & $63$  & $12.7$ \\
\quad infant mortality (per 1000)    & $199$ & $\mathbf{7.90}$ & $\mathbf{244}$ & $\mathbf{49}$ \\
\quad internet users (\% of pop.)    & $183$ & $0.529$ & $16$  & $3.3$  \\
\quad life expectancy (years)        & $220$ & $0.263$ & $8.1$ & $1.6$  \\
\quad school life expectancy (years)$^{\dagger}$ & $166$ & $0.268$ & $8.3$ & $1.7$  \\
\bottomrule
\end{tabular}
\caption{Variance of $\ln \rho_{\text{eff}}$ across alternative outcomes.
Source: \protect\path{analysis/output/multi_outcome.json}.
$^{\ddagger}$The 6 variants are gender breakdowns of the race-pooled
outcome (\texttt{\_pooled\_male\_}, \texttt{\_pooled\_female\_},
\texttt{\_pooled\_pooled\_}) and race breakdowns of the gender-pooled
outcome (\texttt{\_white\_pooled\_}, \texttt{\_black\_pooled\_},
\texttt{\_hisp\_pooled\_}); they are not the full $3{\times}2$ Cartesian
race$\times$gender grid, only the marginal univariate breakdowns.
$^{\dagger}$School life expectancy ($\texttt{SE.SCH.LIFE}$) is reported
to the World Bank with a multi-year lag; the API has \emph{no} 2022
values at extraction time (April~2026). For each country we therefore
use the latest available year (mostly $2018$--$2019$, range
$2010$--$2019$), giving $n=166$ countries after the WB-aggregate
filter. The reading is therefore ''most recent reported value per
country'' rather than a strict 2022 cross-section, and the figure
should be read as informative but not strictly comparable to the other
2022 rows of the table.}
\label{tab:multi-outcome}
\end{table}

\begin{figure}[h]\centering
\includegraphics[width=.95\linewidth]{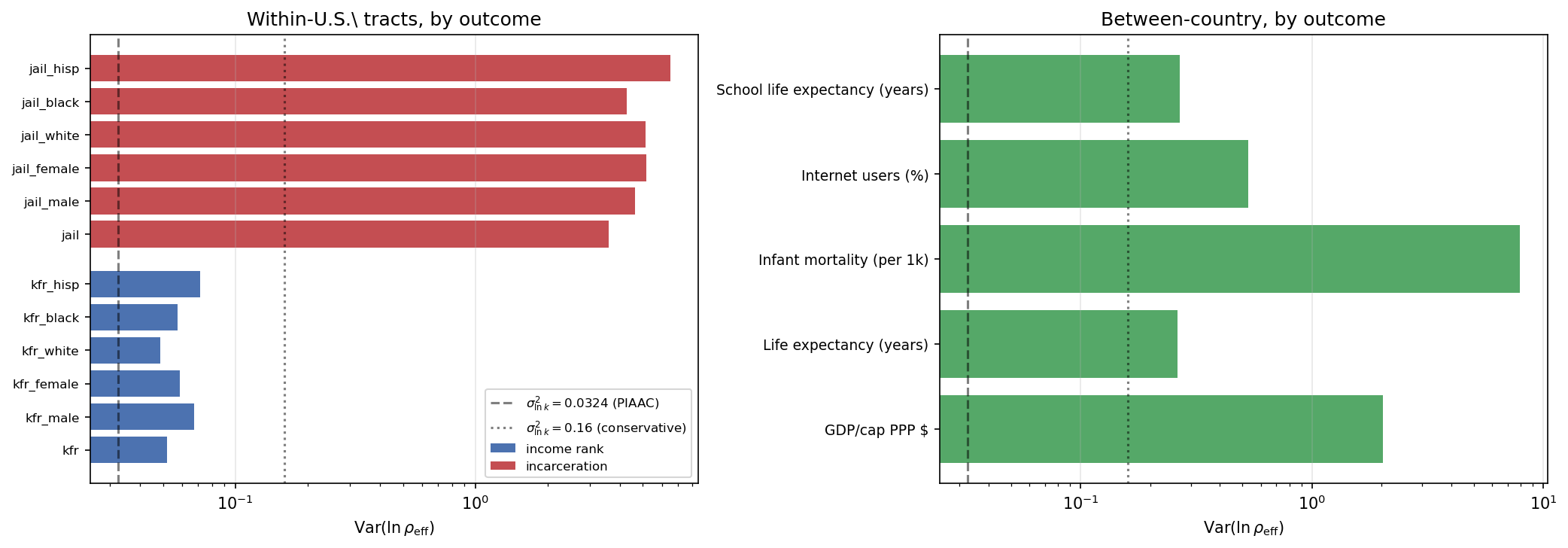}
\caption{Variance of $\ln \rho_{\text{eff}}$ across alternative outcomes
(log $x$-axis). Vertical lines mark $\sigma_{\ln k}^2 \in \{0.0324, 0.16\}$.}
\label{fig:multi-outcome}
\end{figure}

The dominance inequality is outcome-dependent. For income (within-country
deciles and between-country GDP), infant mortality (between-country
$R \approx 244$ at PIAAC $\sigma$), and incarceration ($R$ above $100$
even within a single census tract), the per-tract or per-country
dispersion is large enough that any plausible capability dispersion is
dominated. For life expectancy ($n=220$) the picture is weaker but still dominant:
median country values now sit in a narrow band ($70$--$85$ years), and
at the conservative bound $\sigma_{\ln k} = 0.40$ the implied $R$ falls
to $1.6$, still above unity, but no longer by orders of magnitude.
School life expectancy points the same way ($R \approx 1.7$ at the
conservative bound, $n=166$ on the latest available year per country,
mostly $2018$--$2019$): the international band has narrowed enough
that the dominance margin shrinks to a single-digit factor, mirroring
the life-expectancy result. Life expectancy is the outcome where the
dominance margin is thin enough that capability dispersion is
comparable to environmental dispersion.

The original abstract claim of~\cite{sestak2026env} thus needs to be
read with an outcome qualifier. The inequality
$\operatorname{Var}(\ln \rho_{\text{eff}}) \gg \operatorname{Var}(\ln
k)$ is robust for economic and life-or-death outcomes; it shrinks
toward parity for outcomes that have been the target of sustained
global convergence (life expectancy as the robust example). The dichotomy is
informative on its own: the outcomes whose dominance margin shrinks
to a single-digit factor are exactly those where international policy
effort has been concentrated for the last fifty years.

\section{Discussion}

\subsection{Confirmation, with refinements}

The empirical decomposition confirms the prediction~\eqref{eq:claim} at
the global, between-country and within-country-decile levels, with a
dominance ratio $R \in [27, 134]$ at the global level. Four refinements
to the original statement emerge.

\textbf{Magnitude.} ''Two to three orders of magnitude'' in the original
abstract should be read as one to two orders.

\textbf{Scope.} The result holds for environment in the broad sense
(geography, class and family taken together). It does not hold within
already-homogenized sub-units such as a single U.S.\ census tract for
income outcomes; the within-tract regime is the locally meritocratic
one.

\textbf{Composition.} The dominant component is within-country
inter-class dispersion ($3.29$), not between-country dispersion ($1.04$).
The geographic framing of the original paper (''Silicon Valley vs.\
remote village'') is empirically the smaller of the two dominant axes,
and the relative weight of the two has shifted further toward class over
the past three decades (Table~\ref{tab:timeseries}).

\textbf{Outcome dependence.} The dominance inequality is robust for
income, infant mortality and incarceration, but weaker for outcomes that
have undergone sustained global convergence (life expectancy).
The original framework treats ''success'' as a single binary event; the
data show that the magnitude of $R$ depends on which success criterion
one chooses, and that two distinct outcome classes emerge (those where
international policy has homogenized environments, and those where it
has not).

\subsection{Partial identification of \texorpdfstring{$\rho_{\text{eff}}$}{rho\_eff} versus $k$}

The decomposition above measures the variance of $\ln$ income across
atoms, which is the variance of the outcome, not of $\rho_{\text{eff}}$
alone. The outcome combines $\rho_{\text{eff}}$ and $k$ via the
model~\eqref{eq:model}, so the measured $4.33$ is an upper bound on
$\operatorname{Var}(\ln \rho_{\text{eff}})$. Let $\alpha \in [0, 1]$ be
the share of the within-country decile dispersion ($3.29$) that is
''really'' capability rather than environment. If the literature-derived
$\sigma_{\ln k}^2$ is taken as a baseline that does not yet include this
share, the reattribution moves both the numerator and the denominator:
\begin{equation}\label{eq:radj}
    R^{\text{adj}}(\alpha) =
    \frac{\operatorname{Var}(\ln \rho_{\text{eff}}) - \alpha \cdot 3.29}
         {\sigma_{\ln k}^2 + \alpha \cdot 3.29}
    = \frac{4.33 - \alpha \cdot 3.29}{\sigma_{\ln k}^2 + \alpha \cdot 3.29}.
\end{equation}
The denominator inflation is essential: if half of the within-country
dispersion is ''really'' capability, then the implied $\sigma_{\ln k}^2$ is
no longer $0.16$ but $0.16 + 1.645 \approx 1.81$, and the ratio falls
sharply.

\begin{table}[h]\centering
\begin{tabular}{rcc}
\toprule
$\alpha$ (share of within-country attributed to $k$) & $R^{\text{adj}}$ at
$\sigma_{\ln k}=0.18$ & $R^{\text{adj}}$ at $\sigma_{\ln k}=0.40$ \\
\midrule
$0.00$  & $133.7$ & $27.1$ \\
$0.05$  & $21.2$  & $12.8$ \\
$0.10$  & $11.1$  & $8.2$  \\
$0.20$  & $5.3$   & $4.5$  \\
$0.50$  & $1.6$   & $1.5$  \\
$1.00$  & $0.31$  & $0.30$ \\
\bottomrule
\end{tabular}
\caption{Adjusted dominance ratio $R^{\text{adj}}(\alpha)$ from~\eqref{eq:radj}.
Reattribution is more punishing than the simple-numerator-only formula would
suggest because the implied $\sigma_{\ln k}$ also inflates.}
\label{tab:radj}
\end{table}

The dominance result therefore survives modest reattribution
($\alpha \lesssim 0.1$ leaves $R$ around an order of magnitude above
unity) but not aggressive reattribution: $\alpha \approx 0.5$ collapses
$R$ to roughly $1.5$, and $\alpha = 1$ inverts the conclusion. The
literature on capability dispersion does not support the aggressive
case. With $\alpha = 0.5$ the implied $\sigma_{\ln k}$ would be near
$1.34$, corresponding to a 95/5 percentile spread of cognitive ability
of about $80$-fold; the largest published estimates (Schmidt--Hunter,
high-complexity tasks) give a spread on the order of $3$-fold. The
conservatively defensible regime is therefore $\alpha \lesssim 0.1$, in
which $R$ remains in $[8, 21]$. A proper resolution of the
partial-identification problem would require a causal-mobility
design~\cite{chetty2018atlas} that we leave for future work.

\subsection{Selection bias}

If more capable individuals systematically sort into better
environments, the covariance
$\operatorname{Cov}(\ln \rho_{\text{eff}}, \ln k)$ is positive and the
decomposition above attributes part of the sorting effect to environment
rather than to capability. Resolving this cleanly would require a
causal-mobility design such as the ''movers'' identification of
Chetty et al.~\cite{chetty2018atlas}; we cannot replicate it here
without access to their linked tax-and-census micro-data, but the
published bounds let us put a number on the worst-case impact.

Chetty et al. estimate that approximately $\mathbf{60\%}$ (CI roughly
$50$--$70\%$) of the cross-tract variation in children's adult-income
rank for fixed-percentile parents is a \emph{causal place effect};
the residual $\mathbf{40\%}$ ($30$--$50\%$) is family / individual
selection into neighborhoods. Treating this as an upper bound on the
selection share of \emph{any} of our environment-attributed dispersion
levels (a deliberately pessimistic move, because cross-country
dispersion is implausibly more selection-driven than cross-tract
dispersion within a metropolitan labor market) gives the
selection-corrected dominance ratios in
Table~\ref{tab:selection-bounds}.

\begin{table}[h]\centering
\begin{tabular}{lccc}
\toprule
$\sigma_{\ln k}$ & no correction & $40\%$ selection & $50\%$ selection \\
& ($R$) & ($R^{\text{sel}}_{60}$) & ($R^{\text{sel}}_{50}$) \\
\midrule
$0.18$ (PIAAC)        & $\mathbf{134}$ & $\mathbf{80}$ & $\mathbf{67}$ \\
$0.40$ (conservative) & $\mathbf{27}$  & $\mathbf{16}$ & $\mathbf{14}$ \\
\bottomrule
\end{tabular}
\caption{Dominance ratio after subtracting a Chetty-style selection
share from the global pooled $\operatorname{Var}(\ln \rho_{\text{eff}})
= 4.33$. Even at the upper end of the published causal bounds (only
half of cross-tract variation is genuinely environmental), the
inequality $R \gg 1$ survives by an order of magnitude.}
\label{tab:selection-bounds}
\end{table}

The reason the order-of-magnitude conclusion is robust to selection is
geometric: for $R$ to drop below unity, selection would have to
account for $\geq \tfrac{4.33 - \sigma_{\ln k}^2}{4.33}$ of the
measured variance, i.e.\ $\geq 99.3\%$ at PIAAC $\sigma$ and $\geq
96.3\%$ at the conservative bound. Published causal estimates leave at
most half of the variation as selection, so the residual that survives
the strongest defensible correction is still in $[14, 80]$ depending
on $\sigma_{\ln k}$. A within-country causal-mobility analysis (the
WID atoms here are not directly compatible with the Opportunity Atlas
panel, but the same identification could in principle be run on
WID-pooled cross-country movers from the Lakner--Milanovic survey
data~\cite{lakner2016}) is left for future work.

\subsection{Interpretation of luck}

The original paper~\cite{sestak2026env} positioned its framework against
the agent-based simulation of Pluchino et al.~\cite{pluchino2018}, which
talks about ''luck'' where we talk about ''environment''. The two are
not in opposition. ''Luck'' is just the residual of $k \ll n$: with $k$
realized attempts and an unbounded number of unmodeled co-determinants,
each attempt is stochastic relative to the model. Our decomposition
measures the dispersion of the atom-level success rate
$\rho_{\text{eff}}$ across (country, decile) atoms and is silent on the
within-atom stochasticity that produces Pluchino's Pareto wealth tails.
Pluchino populates the within-atom draws; we populate the between-atom
dispersion of the rate against which those draws are made.

\paragraph{Operational status of an ''attempt''.} It is worth being
explicit about how an attempt is to be read in the model, because the
notation $k$ might suggest a count of identical Bernoulli trials.
Attempts in the model of~\cite{sestak2026env} are not identical and not
repeatable. Each attempt is anchored to a specific point in time and
space: it occurs only once, in a configuration of the world that will
not recur. Two consequences follow.

First, the attempt is jointly determined by the entity and the
environment, and \emph{always} by both. The entity does not act on a
passive backdrop; the entity itself changes between attempts (skills
acquired, beliefs updated, body and capital aged, social position
shifted), and the environment changes too (other agents act, prices
move, opportunities open and close). There is no ''ceteris paribus''
draw in the actual sequence of a life; each $i$-th attempt is
embedded in a unique context $(t_i, x_i)$ that determines what the
relevant possibility space $A(E,P)$ even is. The treatment of
$\rho_{\text{eff}}$ as a fixed atom-level rate in the present
decomposition is therefore an aggregation: we average the per-attempt
$\rho_{\text{eff},i}$ over the attempts available within an atom
(country, decile, tract).

Second, this is the deeper reason ''luck'' and ''environment'' are not
rival explanations. Both names point at the same fact: that the
entity does not control the configuration in which its attempt occurs.
Pluchino models the per-attempt randomness of the configuration directly
(stochastic encounters with ''lucky'' or ''unlucky'' events); we
collapse the same randomness into the dispersion of the configuration
rate $\rho_{\text{eff}}$ across atoms. The two are different
levels-of-detail descriptions of the same physical fact: the attempt is
unique, time- and space-bound, jointly co-produced by the entity
(itself changing) and external processes, and never re-played.
Capability $k$ is the count of such unique events within a finite life;
it bounds how many configurations the entity can sample, but does not
control which configurations those are.

\section{Conclusion}

The structural inequality $\operatorname{Var}(\ln \rho_{\text{eff}}) \gg
\operatorname{Var}(\ln k)$ proposed in~\cite{sestak2026env} is
empirically confirmed at the global pooled-individual level for income,
with a dominance ratio between 27 and 134 across plausible calibrations
of capability dispersion. The confirmation survives bootstrap
resampling at the within-tract level (95\% CI $[0.050, 0.054]$ on
$\operatorname{Var}(\ln \rho_{\text{eff}})$), the choice of success
threshold and per-life capacity $\bar k$, and modest reattribution of within-country dispersion to capability
($\alpha \lesssim 0.1$ leaves $R$ above 8). The margin is not uniform
across outcomes: for life expectancy, international convergence has
narrowed country bands enough that $R$ falls to a single-digit factor
($1.6$--$8.1$) rather than orders of magnitude, and the within-tract
scope is similarly outcome-dependent: income at the within-tract
level gives $R \in [0.33, 1.61]$, straddling parity between capability
and environment, whereas incarceration at the same spatial scale gives
$R$ between $22$ and $201$ depending on $\sigma_{\ln k}$, so
environment still dominates by one to two orders even within a single
tract. Nor is the margin uniform across aggregation levels: at the
between-country level (WID decomposition) $R$ ranges from $32$ at PIAAC
$\sigma$ down to $6.5$ at the conservative bound, so the
''one-to-two-orders'' headline applies to the global pooled-individual
and within-country-decile levels but only marginally to the
between-country component.

What survives is therefore narrower and more interesting than the
original universal claim. ''Where you are born, into which class, into
which family'' matters by one to two orders of magnitude more than
''how capable you are'' for the outcomes that determine economic
position and life-or-death circumstances; for the outcomes that
international policy has spent half a century equalizing (life
expectancy in non-failing states), the gap narrows
from orders of magnitude to a single-digit factor; environment
still dominates, but capability becomes a comparable input. The
meritocratic intuition is correct inside a homogenized sub-population
(within a single census tract for income), and partial for outcomes
that have already been homogenized across populations. For everything
else, environment dominates by one to two orders.

\section*{Reproducibility}

The full analysis pipeline (raw data downloads, decomposition notebooks,
figures, and this paper's tables) is shipped alongside this document in
the \texttt{analysis/} directory and is also available as a public
repository at
\href{https://github.com/dkrse/environment-dominance}{\texttt{github.com/dkrse/environment-dominance}}.
Each input file is hashed in
\path{analysis/data/MANIFEST.md}; each output artefact is hashed in the
append-only \path{analysis/output/RESULTS_MANIFEST.md}; integrity can be
verified at any time by calling \texttt{verify()} in
\path{analysis/notebooks/_results_io.py}. Re-running the notebooks via
\texttt{jupyter nbconvert} reproduces every headline number used in
Tables~\ref{tab:decomp}--\ref{tab:multi-outcome} from the cached
datasets in under a minute on a 2024-class laptop. The two derived
tables (Table~\ref{tab:radj} on partial-identification reattribution
and Table~\ref{tab:selection-bounds} on selection-bias bounds) are
computed from the same headline values $4.33$ and $3.29$ via the
formulas given in the text.

\section*{Data availability}
Three public datasets, all included in the accompanying
\texttt{analysis/data/} directory with provenance and SHA-256 hashes:
Opportunity Atlas (Opportunity Insights, October 2018 release); World Bank
Open Data API (\texttt{NY.GDP.PCAP.PP.KD}, \texttt{SP.POP.TOTL}, 2022); World
Inequality Database bulk download (variable \texttt{sptincj992}, April 2026
release).

\section*{Acknowledgements}
The author received no specific funding for this work and declares no
competing interests.



\begin{thebibliography}{9}

\bibitem{sestak2026env} Sestak, K. (2026). The Dominance of Environment over
Entity's Capabilities. \textit{arXiv preprint} arXiv:2605.02985, submitted
4~May 2026. \url{https://arxiv.org/abs/2605.02985}.

\bibitem{chetty2014} Chetty, R., Hendren, N., Kline, P., Saez, E. (2014). Where
is the Land of Opportunity? The Geography of Intergenerational Mobility in the
United States. \textit{Quarterly Journal of Economics}, 129(4), 1553--1623.
\href{https://doi.org/10.1093/qje/qju022}{doi:10.1093/qje/qju022}.

\bibitem{chetty2018atlas} Chetty, R., Friedman, J.\,N., Hendren, N., Jones,
M.\,R., Porter, S.\,R. (2018). The Opportunity Atlas: Mapping the Childhood
Roots of Social Mobility. \textit{NBER Working Paper} 25147.
\href{https://doi.org/10.3386/w25147}{doi:10.3386/w25147}.

\bibitem{milanovic2015} Milanovic, B. (2015). Global Inequality of Opportunity:
How Much of Our Income Is Determined by Where We Live? \textit{Review of
Economics and Statistics}, 97(2), 452--460.
\href{https://doi.org/10.1162/REST_a_00432}{doi:10.1162/REST\_a\_00432}.

\bibitem{lakner2016} Lakner, C., Milanovic, B. (2016). Global Income
Distribution: From the Fall of the Berlin Wall to the Great Recession.
\textit{World Bank Economic Review}, 30(2), 203--232.
\href{https://doi.org/10.1093/wber/lhv039}{doi:10.1093/wber/lhv039}.

\bibitem{piketty2014} Piketty, T. (2014). \textit{Capital in the
Twenty-First Century}. Harvard University Press. ISBN 978-0-674-43000-6.
\href{https://doi.org/10.4159/9780674369542}{doi:10.4159/9780674369542}.

\bibitem{pluchino2018} Pluchino, A., Biondo, A.\,E., Rapisarda, A. (2018).
Talent vs Luck: The Role of Randomness in Success and Failure.
\textit{Advances in Complex Systems}, 21(3--4), 1850014.
\href{https://doi.org/10.1142/S0219525918500145}{doi:10.1142/S0219525918500145}.

\bibitem{schmidt1998} Schmidt, F.\,L., Hunter, J.\,E. (1998). The Validity and
Utility of Selection Methods in Personnel Psychology: Practical and
Theoretical Implications of 85 Years of Research Findings. \textit{Psychological
Bulletin}, 124(2), 262--274.
\href{https://doi.org/10.1037/0033-2909.124.2.262}{doi:10.1037/0033-2909.124.2.262}.

\bibitem{wid2026} Alvaredo, F., Chancel, L., Piketty, T., Saez, E., Zucman, G.,
et al. (2026). \textit{World Inequality Database (WID.world)}, bulk download
(April 2026 release). \url{https://wid.world/}.

\bibitem{worldbank2022} World Bank (2022). \textit{World Development
Indicators}: \texttt{NY.GDP.PCAP.PP.KD}, \texttt{SP.POP.TOTL},
\texttt{SI.POV.GINI}. \url{https://data.worldbank.org/}.

\bibitem{piaac} OECD (2013, 2019). \textit{Programme for the International
Assessment of Adult Competencies (PIAAC), Survey of Adult Skills}.
\url{https://www.oecd.org/skills/piaac/}.

\bibitem{hanushekwoessmann} Hanushek, E.\,A., Woessmann, L. (2012). Do Better
Schools Lead to More Growth? Cognitive Skills, Economic Outcomes, and
Causation. \textit{Journal of Economic Growth}, 17(4), 267--321.
\href{https://doi.org/10.1007/s10887-012-9081-x}{doi:10.1007/s10887-012-9081-x}.

\end{thebibliography}
\end{document}